# Depth-adapted adaptive optics for three-photon microscopy

**QI HU,[1,†] JINGYU WANG,[1,†] HURIYE ATILGAN,[2,†] ARMIN LAK,[2] AND MARTIN J. BOOTH,[1,*]**

[1]*Department of Engineering Science, University of Oxford, Oxford, UK*
[2]*Department of Physiology, Anatomy, and Genetics, University of Oxford, Oxford, UK*
[†]*These authors contributed equally.*
* *martin.booth@eng.ox.ac.uk*

**Abstract:** Three-photon (3-P) fluorescence microscopy enables deep in vivo imaging with subcellular resolution, but its performance is fundamentally constrained by the maximum permissible laser power required to avoid tissue heating and photodamage. Under these power-limited conditions, fluorescence signal generation, image contrast, and achievable imaging depth are strongly affected by the illumination beam profile and aberration correction strategy. In this paper, we showed that using a fixed illumination beam size was suboptimal across different imaging depths. We further showed that conventional Zernike-based adaptive optics (AO) correction degrades under reduced Gaussian illumination beam sizes due to loss of modal orthogonality. This degradation results in slow convergence, unintended focal and field-of-view shifts, and excessive wavefront deformations. To overcome these limitations, we introduced a depth-adapted AO framework in which both the illumination beam profile and the aberration correction basis were dynamically matched to the imaging conditions. By combining depth-optimised beam underfilling with a bespoke set of illumination-matched aberration modes, we achieved faster and more stable AO convergence, enhanced fluorescence signal and image quality during deep in vivo multi-channel neuroimaging. Together, these results established a practical and robust AO-enabled three-photon microscopy strategy that maximised imaging performance under realistic power constraints.

## 1. Introduction

Three-photon (3-P) microscopy, for its long excitation wavelength and non-linear absorption, has become one of the most promising less-intrusive in-vivo deep imaging techniques for both structural and functional studies [1-5]. In multiphoton imaging, the non-linear absorption ensures intrinsic optical sectioning, i.e. fluorescence occurs only at the focal volume. Compared to two-photon imaging, the longer wavelength light is less susceptible to scattering and allows the illumination to penetrate deeply into the tissue; the cubic dependence on excitation intensity also produces a superior signal-to-background ratio. In contrast to other deep imaging techniques such as computed tomography (CT), magnetic resonance imaging (MRI) or GRIN lens imaging, multi-photon microscopy remains a less invasive imaging modality that offers subcellular spatial resolution. Combined with sparse or cell-type–specific labelling, the technique becomes highly suitable for functional imaging of dynamic biological processes, including the monitoring of calcium transients using genetically encoded indicators such as GCaMP [6, 7].

Despite these advantages, there are still challenges in 3-P imaging. As the illumination beam propagates deeper into the tissue, the absorption and any residual scattering significantly reduce the signal-to-background for deep focusing. Illumination power exponentially decreases with the propagation distance due to dissipation, while energy absorption increases with prolonged exposure; this could lead to phototoxicity, physiological changes, and photothermal damage if the illumination power is not carefully selected, particularly where short pulsed lasers are used [7-9]. Therefore, optimising illumination to maximise fluorescence yield within the limits imposed by thermal damage is critical for extending the imaging capability, especially for animal imaging.

One proposed strategy to overcome these limitations is to adjust the spatial profile of the excitation beam by optimising the underfilling factor at the objective back aperture. Simulation studies have suggested that adjusting the effective illumination numerical aperture (NA) in this way can influence the trade-off between multiphoton fluorescence excitation efficiency and light attenuation [10]. A larger NA theoretically produces a tighter focal volume, but at the cost of a longer optical path due to higher propagation angles through tissue, leading to higher overall absorption. These competing effects imply that the optimal underfilling ratio depends on the imaging depth and the tissue's optical properties. It is well known that an underfilled aperture is preferable for illumination, but a high-NA objective remains advantageous for fluorescence collection. Although several studies have implemented underfilling in multiphoton-photon microscopy, systematic experimental demonstrations across different tissue depths are limited in scope [11].

Despite the long wavelength illumination being less susceptible to tissue scattering, aberration correction using adaptive optics (AO) remains crucial for restoring imaging quality in multi-photon microscopes [12-15]. AO compensates wavefront distortions arising from optical misalignment, sample mounting, surgically implanted optical interfaces (e.g., cranial windows), and structural inhomogeneities within the tissue, thereby significantly improving imaging performance [16]. In practice, many multiphoton imaging systems employ sensorless AO, in which aberrations are inferred indirectly from image-based metrics rather than measured with a dedicated wavefront sensor. Conventional sensorless methods often rely on Zernike polynomial modes for wavefront correction. A major reason for this is the modes' orthogonality over uniformly illuminated pupils. However, their effectiveness is compromised when used in systems with underfilled illumination, which commonly produces a truncated Gaussian beam profile.

Zernike polynomials have sharp peripheral slopes; this effectively means that most modulation power concentrates close to the edges of the AO active area, where little laser energy is present due to the Gaussian beam profile. The mismatch between the Gaussian profile and the ideal uniform illumination affects orthogonality between the modes and results in slow and ineffective aberration compensation. In addition, to correct aberrations for Gaussian beams using Zernike modes where central gradients are moderate, it normally requires modes with huge coefficients for effective correction. Large modal coefficients lead to sharp corresponding peripheral gradients, which place difficult demands on AO correction elements. Although some modified Zernike polynomials have been derived that achieve perfect orthogonality under Gaussian weighting, such modes tend to have even sharper peripheral slopes, creating further difficulties for practical application [17]. This highlights the need for an alternative modal basis that is both orthogonal under Gaussian illumination and practical to implement experimentally.

In this paper, we present a novel adaptive 3-P microscopy scheme, whereby illumination profiles and aberration basis modes are adapted according to the imaging depth. This scheme is demonstrated through practical imaging in awake mouse brain cortex. We undertook a systematic experimental comparison of different illumination configurations and their effects on the imaging quality under constant total laser power at different imaging depths. We introduced a new set of orthogonal modes – weighted-Bessel modes – which are both theoretically orthogonal and practical to use for Gaussian illumination. We compared sensorless AO performance using conventional Zernike modes and weighted-Bessel modes for correcting sample induced aberrations when imaging deep inside awake mouse brains. Weighted-Bessel modes optimised for a specific imaging depth yielded more robust correction, with faster and more stable convergence, achieving superior results in deep imaging.

## 2. Methods

### *2.1 Three-photon adaptive optics microscope*

3-P fluorescence imaging was performed with a Cronus 3P system (Light Conversion, Lithuania) optimised for 1300 nm excitation (Figure 1a, See supplementary information for details of the system). Briefly, imaging was conducted using a galvo scanner-based multiphoton microscope with a 25× water immersion objective (1.0 NA, 4 mm WD). The tunable beam expander (Thorlabs ZBE1C) was positioned before the deformable mirror so that the conjugation and magnification between the deformable mirror, scanners, and pupil back aperture was not affected when the beam size was adjusted (Figure 1a). By tuning the beam expansion, we controlled the degree of objective back-aperture filling, allowing a trade-off between focal confinement and excitation path length in tissue (Figure 1b).

The illumination beam size at the objective back aperture was parameterised by the underfilling ratio β, defined as the ratio of the Gaussian beam diameter ($1/e^2$) to the objective back-aperture diameter. A beam expander allowed us to optimise the Gaussian beam profile with a continuously tuneable zoom factor of 0.5× to 2.5×, corresponding to β values ranging from 0.45 to 0.85.

The limiting aperture of the optical system was the deformable mirror (DM) aperture. Dependent on the particular experiment, we employed as the DM either a Imagine Optic Mirao-52e (diameter 15 mm) or an Alpao DM97 (diameter 13.5 mm). These were mapped respectively to 85% and 77% (in diameter) of the objective back aperture. This system configuration was designed to allow a good use of DM actuator strokes when a reduced illumination beam size was used for imaging. From the simulation results (Figure 1c), the optimised underfilling ratio β for any imaging depths more than 600 μm was always below 0.75, smaller than the DM aperture size projected at the objective pupil, showing that our choice of magnification is suitable. Any further beam expansion would not lead to a higher filling ratio, as it was imposed by the limit of the DM aperture, but only produce a flattened beam profile.

### *2.2 Illumination beam size optimisation*

Two opposing factors influence how a beam travels through tissue and how much 3-P signal it generates. A high-NA system focuses light more tightly to form a smaller focal volume and enhance 3-P fluorescence generation. However, the wider illumination angle leads to longer propagation paths through tissue, causing greater power loss from absorption and scattering. Using the numerical method of [10], we optimised how much the objective's back aperture should be underfilled by a Gaussian beam in order to maximise fluorescence at different imaging depths in tissue (see the supplementary information for more details). Using the results of the attenuation depth measurement of different brain structures in literature [10, 18], we optimized the beam profile expansion for different imaging events numerically, and we verified performance experimentally. For experimental verification, we calibrated the optical system for accurate adjustment of the illumination beam expansion (see details of calibration in Supplementary info). The average laser power, measured after the objective lens, was adjusted to maintain constant when changing the beam size for fair comparisons.

### *2.3 Weighted-Bessel (w-Bessel) Mode*

Many implementations of AO in microscopy use indirect wavefront sensing (sensorless) methods to correct aberrations with AO devices. Such methods normally compensate aberrations iteratively using an orthogonal set of mode bases, deriving the optimised coefficient of each modal component.

There is an inherent issue with Zernike modes when used to correct aberrations in a (truncated) Gaussian beam. The beam profile means that the orthogonality of Zernike polynomials is degraded, leading to modal cross-coupling and unstable convergence (details of this analysis are discussed in the results section). To overcome these limitations, we derived a new set of Bessel function-based modes, called weighted-Bessel (w-Bessel) modes, that are orthogonal for Gaussian beam of different sizes. Similar to Zernike modes $Z_n^m$, the mode $B_n^m$ has a radial order of $n$ and an angular order of $m$, where $|m| \leq n$ and $m = n - 2l$ such that $l$ is a whole number. This should not be confused with "Bessel beams" that have a Bessel amplitude function and have also been used in 3-P imaging for extending the depth of focus (Chen, et al. 2018). We define w-Bessel modes as

$$B_n^m(r,\theta) = N\, J_{|m|}\left(r\, j_{|m|,\frac{n-|m|}{2}+1}\right) \operatorname{ang}(m\theta) \frac{1}{\sqrt{\hat{I}(r,\theta)}} \tag{1}$$

where $N$ is the Zernike normalisation defined as

$$N = \begin{cases} \sqrt{n+1} & \text{if } m = 0 \\ \sqrt{2n+2} & \text{if } m \neq 0 \end{cases}$$

and $\operatorname{ang}(m\theta)$ is the angular term defined the same as for the Zernike polynomials,

$$\operatorname{ang}(m\theta) = \begin{cases} 1 & \text{if } m = 0 \\ \sin(-m\theta) & \text{if } m < 0 \\ \cos(m\theta) & \text{if } m > 0 \end{cases}$$

$J_{|m|}(r)$ is the $|m|$th order Bessel function of the first type and $j_{|m|,\frac{n-|m|}{2}+1}$ is the $\left(\frac{n-|m|}{2}+1\right)$th zero of the $|m|$th order Bessel function of the first type. $\hat{I}(r,\theta)$ is the normalised intensity distribution of the Gaussian illumination profile defined as $\hat{I}(r,\theta) = \frac{I(r,\theta)}{\iint I(r,\theta)\, r\, dr\, d\theta}$. $r$ and $\theta$ are the radial coordinates where $0 \leq r \leq 1$ and $-\pi \leq \theta < \pi$.

The illumination beam profile is included in the w-Bessel mode definition, similar to the approach in the derivation of Gaussian Zernike modes in [17]. W-Bessel modes are thus not a single fixed set of modes but are a class of modes adapted according to the beam expansion, which may be optimised depending on the imaging depth and the attenuation depths of different specimen tissue structures.

These modes were defined heuristically to provide the desired properties of orthogonality and reduced phase magnitude and gradient at the periphery compared to Zernike polynomials. We note that they do not form a complete set; this is readily seen as each of the modes at $r = 1$ is set to zero. However, this is of limited practical concern, as the amplitude of the light reaching the focus at the edge of the aperture is small, hence the phase modulation at the edge has little effect on imaging.

### *2.4 Orthogonality analysis*

To understand the modal independence with respect to a Gaussian illumination profile, we defined the modal orthogonality according to the inner product

$$\langle M_i\,, M_j \rangle = \frac{1}{\pi} \iint M_i\,(r,\theta) M_j(r,\theta)\ \hat{I}(r,\theta)\ r\, dr\, d\theta \tag{1}$$

Where $M_i$ and $M_j$ are the $i$th and $j$th modes of an aberration expansion set; $i$ and $j$ are single indices following (but not limited to) Noll's indices. The orthogonality analysis results of the w-Bessel modes and Zernike modes are computed using the modulus of the inner product and are presented in Figure 2c.

## 3. Results

To optimise the 3-P fluorescence generation, we systematically analysed the effects of the beam size when imaging at different cortical depths in an awake mouse brain. We also compared the performance of the conventional Zernike polynomials with the proposed w-Bessel modes in convergence speed, stability and correction quality when used for compensating sample-induced aberrations through sensorless AO methods.

### *3.1 Depth-dependent adaptation of the illumination beam profile*

We investigated through both modelling and experiment, the trade-offs involved in choosing the optimal objective underfilling ratio for 3-P excitation in different imaging depths, as explained in the method section. We carried out numerical simulations adapted from (Wang, et al. 2015), to analyse the consequences of varying illumination conditions (Figure 1c). In agreement with the previous study, the simulations suggested optimal beam profiles at the objective back aperture to maximise the fluorescence signal at different imaging depths in the mouse cortex.

For experimental validations, the illumination beam size was continuously tunable using the beam expander. To enable a simple comparison fair in average excitation power, we maintained a constant laser power after objective while adjusting the illumination beam size to compare the collected fluorescence intensity. We selected three imaging depths within the medial frontal cortex—600, 1000, and 1300 μm —where the anatomical organisation allows continuous imaging into deeper tissue (see the Beam profile measurement section in supplemental document). At each depth, we imaged the same cortical field of view using different illumination beam sizes. The change in fluorescence intensity with the beam size was consistent with the simulated results (Figure 1d, e).

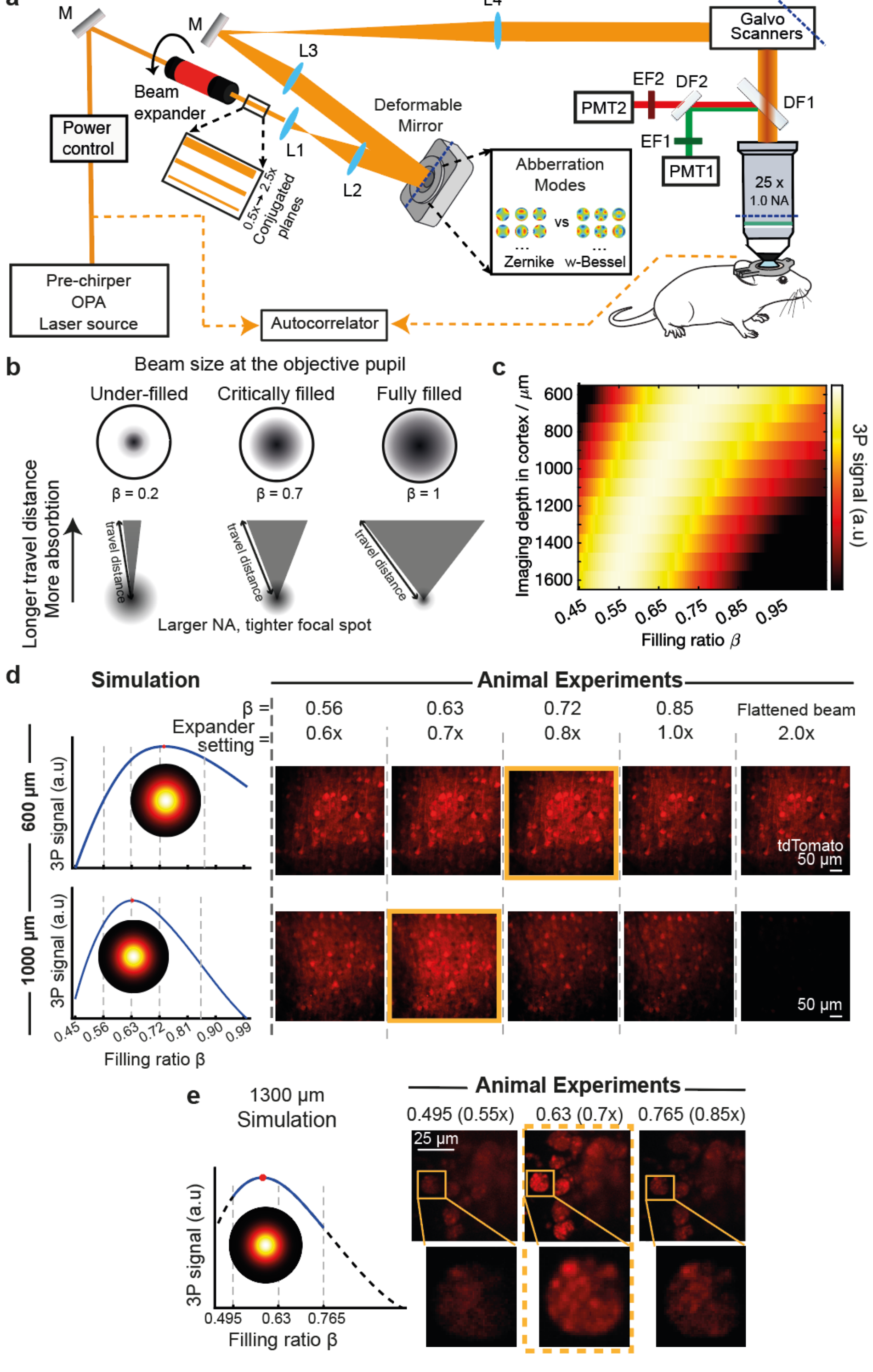

Fig. 1. **Depth-dependent adaptation of the illumination beam profile in three-photon microscopy.** (a) Schematic of the adaptive three-photon microscope, incorporating a beam expander and a deformable mirror, for this study. (b) Conceptual illustration of under-filled, critically filled, and fully filled objective back apertures, illustrating the trade-off between focal spot size, excitation efficiency, and propagation loss as a function of the underfilling parameter β. (c) Numerical simulation predicting the optimal illumination filling ratio at the objective back aperture as a function of imaging depth. At each depth, the simulated fluorescence intensity was normalised such that the maximal intensity equals to 1. (d) In vivo three-photon images with system flat aberration correction applied acquired at 600 and 1000 μm depths in the medial frontal cortex using different illumination beam sizes (β), adjusted via the beam expander. (e) Experimental validation at 1300 μm depth with system flat pre-compensation applied, comparing simulated point-spread functions and corresponding in vivo images for different β values, confirming a further reduction in optimal beam filling at greater depths.

At 600 μm depth, a beam size with the underfilling parameter $\beta \approx 0.72$ produced the brightest images, matching the simulated optimum ($\beta \approx 0.73$) (Figure 1d, top panel). Moving deeper into the tissue, the optimised beam size progressively decreased in simulation, with $\beta \approx 0.63$ at $1000\ \mu m$, $\beta \approx 0.58$ at $1300\ \mu m$ (Figure 1d, e); this trend was verified experimentally. In addition, it was observed that beam size adjustment had a more pronounced impact on image quality at greater depths, as illustrated at $1300\ \mu m$ deep (Figure 1e).

Overall, we confirmed experimentally that the optimal beam expansion systematically decreases with increasing imaging depth, from $\beta \approx 0.73$ at $600\ \mu m$ depth to $0.58$ at $1300\ \mu m$ depth when imaging in the cortex. Our experimental results matched well to the simulations and showed that it was necessary to adapt the beam illumination profile with depth in order to maintain optimum imaging quality.

### 3.2 Weighted-Bessel Mode and Orthogonality Validation

The need for depth adaptation of the illumination profile had consequences for the performance of sensorless AO aberration correction. Due to compromised modal orthogonality under Gaussian underfilling, Zernike-based correction progressively degraded as the beam size decreased. To address the issue, we investigated the properties of the new w-Bessel modes. The orthogonality with respect to the Gaussian intensity distribution was quantified using the inner product defined in Equation 2, providing a robust foundation for evaluating performance against Zernike modes.

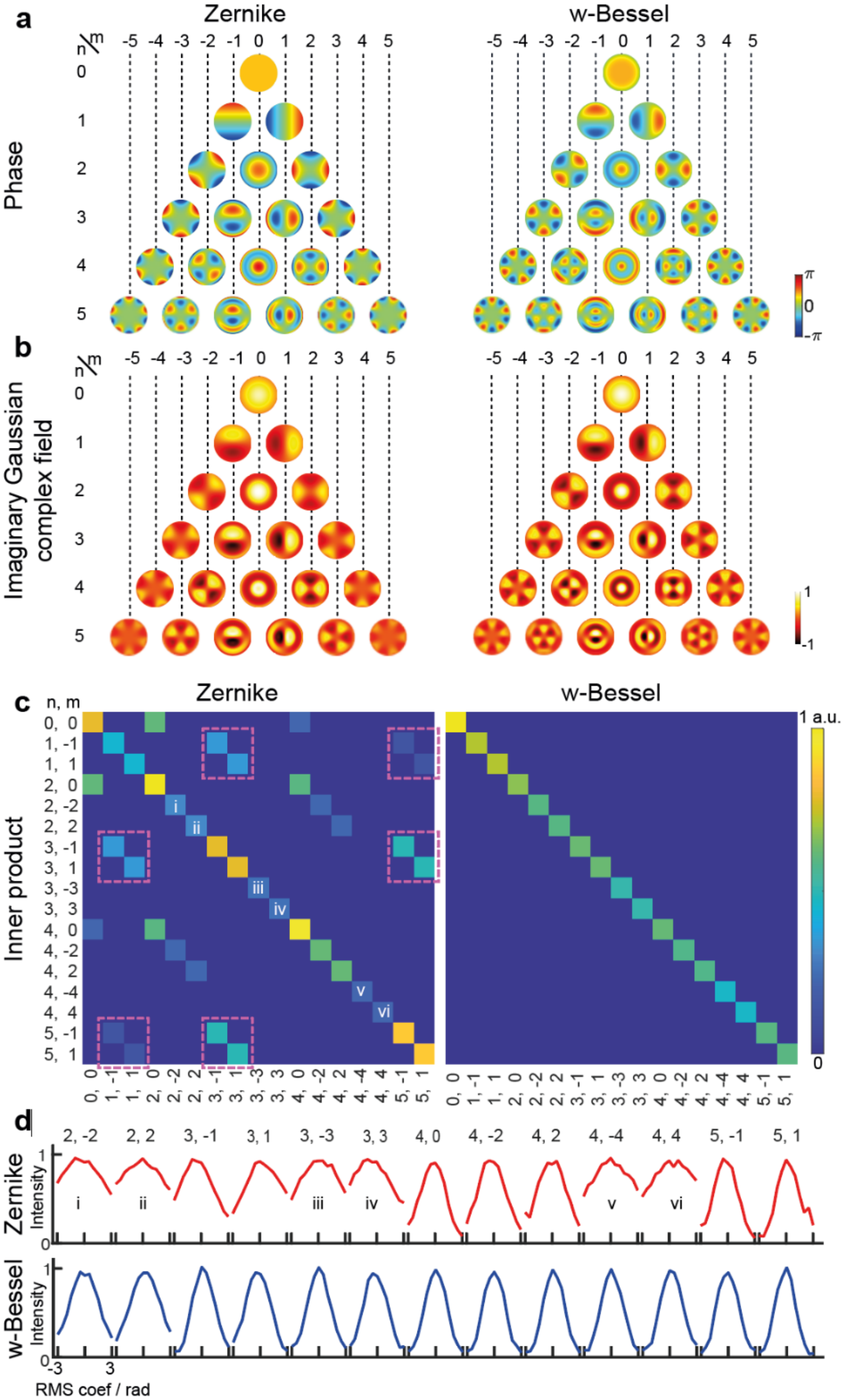

Fig. 2. **Weighted-Bessel modes preserve orthogonality under Gaussian illumination and improve sensorless adaptive optics performance.** (a) Phase patterns of conventional Zernike (left) and weighted-Bessel (w-Bessel; right) modes, shown in the same (n,m) index ordering. (b) Corresponding imaginary part of the Gaussian-weighted complex field for each mode under underfilled illumination. (c) Inner-product matrices (Eq. 2) computed over the Gaussian-weighted pupil, showing increased off-diagonal coupling for Zernike modes and strong diagonal

dominance for w-Bessel modes. (d) Representative sensorless intensity response curves versus applied RMS mode coefficient, illustrating more consistent uni-modal responses for w-Bessel compared to Zernike. Six highlighted high angular order Zernike modes i-vi showed low inner products (in c) and compromised modal influence (in d). for more detailed instructions on labeling supplementary materials in your manuscript. For preprints submitted to Optica Open a link to supplemental material should be included in the submission.

We evaluated the orthogonality of w-Bessel modes and Zernike modes, taking into account the measured Gaussian laser profile ($\beta \approx 0.72$). As shown in Figure 2c, w-Bessel modes are perfectly orthogonal in simulation, indicated by the diagonal matrix; Zernike modes, on the other hand, have crosstalk among modes of the same angular order (m order), as shown by the non-zero off-diagonal components.

While the modal phase patterns differ substantially between bases (Figure 2a), the focal impact of these modes is more clearly understood by examining the resulting complex field of the Gaussian illumination, where both phase and amplitude contribute. For this reason, we compare the imaginary component of the field after applying each mode (Figure 2b). Under Gaussian illumination, the contribution from the pupil periphery is strongly suppressed, leading to similar imaginary field patterns for tip, tilt ($Z_1^{\pm 1}$), coma ($Z_3^{\pm 1}$) and secondary coma ($Z_5^{\pm 1}$) despite their distinct phase profiles. This effect is mitigated for the w-Bessel modes, which better preserve modal differentiation across the illuminated pupil.

We also analysed the intensity variations caused by aberration modes ranging from $-3$ to $+3$ rad rms through experimental measurements on a water-immersed beads sample (Figure 2d). W-Bessel modes showed symmetrical intensity variations of consistent strength for all the selected low-order common modes. In contrast, Zernike modes had inconsistent modulation effects, with high m-order modes introducing only moderate modulations (Figure 2d; astigmatism: $Z_2^{\pm 2}$, trefoil: $Z_3^{\pm 3}$, quatrefoil: $Z_4^{\pm 4}$). This was also reflected in the orthogonality analysis, where the self inner product of the high m-order Zernike modes was small. In contrast, all the modes (apart from the first four low-order modes: $B_0^0$, $B_1^{\pm 1}$, and $B_2^0$) in the w-Bessel set had relatively consistent inner product. (Figure 2c).

Furthermore, the effects of Zernike coma ($Z_3^{\pm 1}$) and its corresponding w-Bessel mode ($B_3^{\pm 1}$) were compared when applied to a bead sample, as shown in Supplementary Figure 4. Results showed that $B_3^{\pm 1}$introduced a larger intensity variation and little change to the centre of gravity of the beads x-y plane. On the other hand, $Z_4^0$ introduced little intensity variation but a large axial shift, which suggested that the spherical had a strong cross-talk with the defocus mode.

Modal orthogonality and field modulation have direct impacts on the convergence speed and correction accuracy when applied to sensorless AO aberration correction. As we will show in the next section, when used for correcting aberrations in deep tissue imaging, Zernike-based correction converged much more slowly, led to extreme resultant peak-to-valley wavefront shapes and compromised imaging performance.

### *3.3 Improved convergence and stability of sensorless AO using weighted-Bessel modes*

To compare the practical performance of Zernike and w-Bessel modes for sensorless AO, we performed in vivo three-photon imaging in an awake cortex of transgenic mice expressing GCaMP in CaMKII-positive neurons, with tdTomato introduced in a subset of cells for dual-colour imaging. The illumination beam size was adjusted according to imaging depth, as determined by the preceding analysis. System aberrations were first corrected using a bead sample sealed by a coverslip of $0.17\ mm$, and the resultant correction was applied as a starting point for all subsequent in vivo experiments.

Imaging was performed at depths of $850\ \mu m$ and $1300\ \mu m$ from the cortical surface (Figure 3). In each case, six low-order modes (astigmatism $Z_2^{\pm 2}$, $B_2^{\pm 2}$, coma $Z_3^{\pm 1}$, $B_3^{\pm 1}$, and trefoil $Z_3^{\pm 3}$, $B_3^{\pm 3}$) were included for correction using either the Zernike or w-Bessel basis. The spherical mode was pre-compensated during the system aberration correction process using both the objective correction collar and the deformable mirror with a bead sample. There was a slight difference in thickness between the sample coverslip ($\sim 0.17\ mm$) and the cranial window ($2 \times 0.1 mm$); the real refractive index of the cortical tissue is often simulated to be 1.36 [3], slightly higher than water, close to 1.32 for $1300\ nm$ light [20]. For this reason and the practically constrained NA, we did not expect, nor did we observe, that additional depth-dependent spherical aberration would have significant impact during animal imaging. Spherical related modes ($Z_4^0$, $B_4^0$) were thus excluded in the animal imaging experiments to ensure a consistent comparison, as both $Z_4^0$ and $B_4^0$, to some different extent, introduce axial focal shift when applied to a truncated Gaussian illumination (see Supplementary Figure 1). There are established practices to remove defocus from spherical modes [21] by modifying the modal definitions. For the scope of this paper, we decided to use the original definitions of Zernike and w-Bessel modes without modifications for a fair comparison.

At $850\ \mu m$ depth, correction using w-Bessel modes converged rapidly, with most modal coefficients approaching zero after the first correction iteration and remaining stable in subsequent iterations (Figure 3c). In contrast, Zernike modes showed substantial coefficient fluctuations across all three iterations, particularly for coma (dashed boxes), indicating unstable convergence. Consistent with this observation, the wavefronts constructed after Zernike correction changed significantly in shape through iterations, but w-Bessel resultant wavefronts retained similar shapes throughout. The final Zernike-corrected wavefront exhibited larger peak-to-valley amplitudes than the w-Bessel solution (Figure 3d).

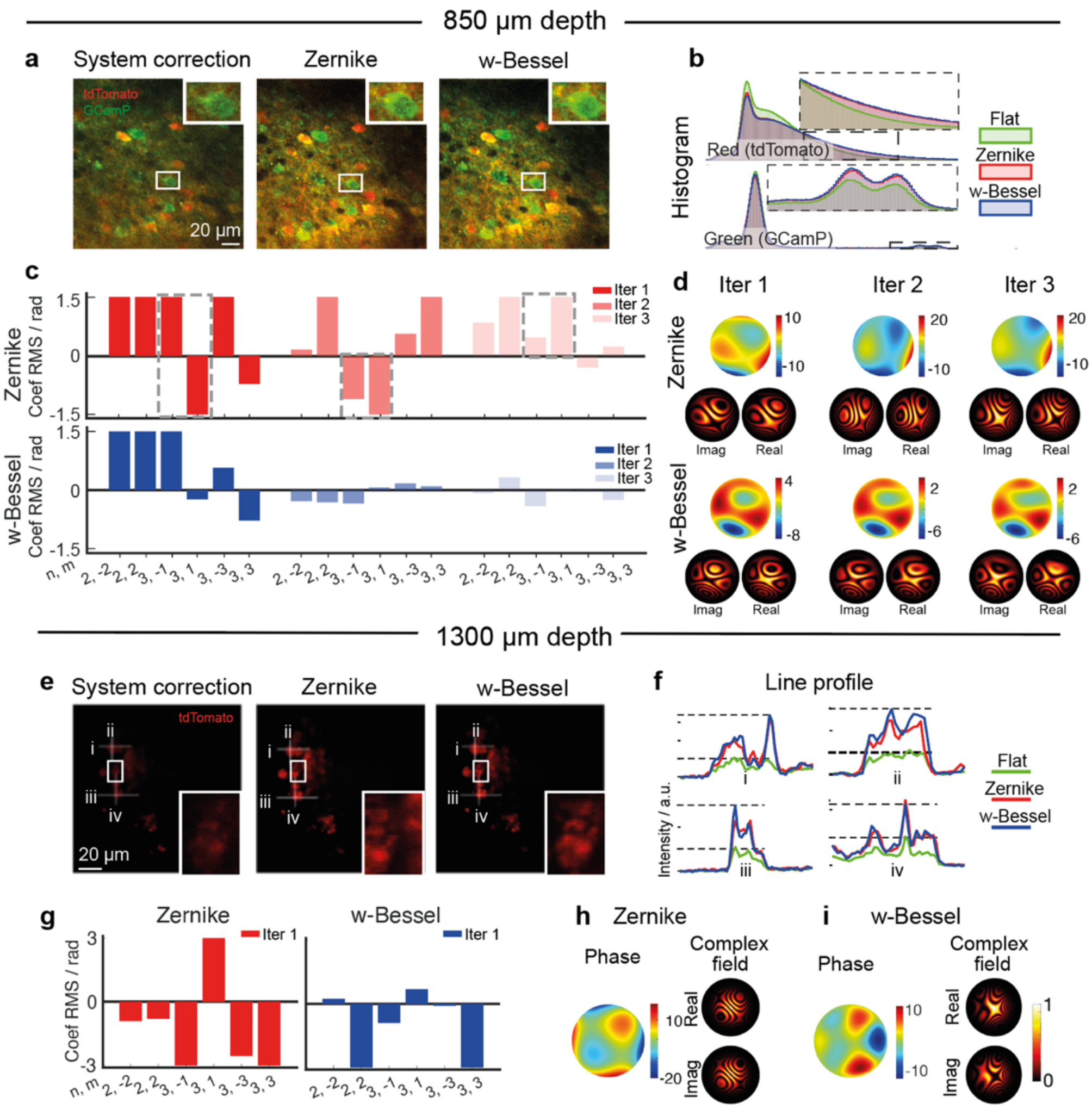

Fig. 3. **Improved convergence, stability, and image quality using weighted-Bessel (w-Bessel) modes for sensorless adaptive optics correction in vivo.** Representative in vivo three-photon images acquired at (a) 850 μm and (e) 1300 μm depth under system-flat conditions, after Zernike- and w-Bessel-based corrections. Insets show magnified regions. Scale bars, 20 μm. (b) Intensity histograms of red (tdTomato) and green (GCaMP) channels for the three correction conditions. Calculated modal coefficients (c) across three successive iterations of sensorless correction at 850 μm depth and (g) after one iteration of correction at 1300 μm depth (d) Phase maps and corresponding complex fields (real and imaginary components) reconstructed from the final wavefront corrections. (f) Line profiles extracted from the regions indicated in (e). (h–i) Phase and complex intensity components of the final Zernike- and w-Bessel-corrected fields, respectively.

Despite these differences in wavefront structure, the reconstructed Gaussian-weighted complex fields showed similar features near the beam centre for both approaches, with discrepancies primarily confined to the pupil edges where the illumination intensity is low. As a result, both correction strategies improved image quality, but w-Bessel modes achieved this improvement using smaller and more stable wavefront corrections, producing brighter and sharper images (Figure 3a, b).

At 1300 $\mu m$ depth, the advantages of w-Bessel modes became more pronounced. Representative images showed stronger signal recovery and improved contrast compared to both system-flat and Zernike-corrected conditions (Figure 3e). Line-profile analysis confirmed increased peak intensity for both correction strategies, with w-Bessel modes providing slightly enhanced contrast and reduced spatial distortions (Figure 3f). Notably, Zernike correction introduced a lateral shift of the field of view due to the introduced coma, whereas w-Bessel correction preserved spatial alignment (Figure 3e, insets).

Overall, w-Bessel modes enabled faster and more stable sensorless AO convergence, thereby requiring fewer acquisitions for effective correction. This improved robustness makes them particularly well-suited for practical in vivo three-photon imaging, where stability and acquisition efficiency are critical. Additional examples from different animals are provided in Supplementary Figure 3.

## 4. Conclusion

Modal-based adaptive optics has traditionally relied on fixed sets of aberration modes applied under fixed illumination conditions, chosen primarily for convenience or historical precedent rather than for optimal performance across imaging depths. In this work, we have addressed the limitations of this approach by introducing an adaptive strategy in which both the illumination beam profile and the aberration correction basis are tailored to the imaging depth and focal position. To our knowledge, this is the first study to theoretically derive and experimentally validate depth-adapted modal bases optimised for deep-tissue three-photon imaging.

Our results demonstrated that adaptable control of the illumination beam profile is essential for efficient fluorescence generation under power-limited conditions. In the mouse cortex, the optimal objective underfilling ratio decreased systematically with depth, shifting from $\beta \approx 0.73$ at 600 $\mu m$ to $\beta \approx 0.58$ at 1300 $\mu m$. This depth-dependent underfilling substantially improved signal generation at depth and highlighted the limitations of fixed illumination strategies commonly used in three-photon microscopy. While we maintained a constant average laser power for a fair comparison, which is reasonable to account for thermal damage, the laser-induced nonlinear tissue damage was ignored. [22] The Gaussian illumination has a higher excitation power density in the centre. The current experiments avoided tissue damage, but we would expect a slightly different relationship if the power density is limited, instead of the total average power, to avoid laser-induced nonlinear damage in deep tissue.

Crucially, this adaptive illumination strategy necessitated a corresponding change in the aberration correction basis. We showed that conventional Zernike modes, while widely used, lose orthogonality under truncated Gaussian illumination, leading to slow convergence, unstable correction and resulting in excessive wavefront deformations and field-of-view shifts. In contrast, the w-Bessel modes introduced here preserve orthogonality under underfilled illumination and enable fast, stable, and reliable sensorless aberration correction. These advantages arise from the intrinsic orthogonality of the w-Bessel basis and are therefore expected to extend beyond the specific indirect AO implementation used in this study. For example, these conclusions are not confined to 3-P microscopy, but would be relevant for any non-linear microscope – including two-photon or harmonic generation – where use of underfilled illumination pupils is advantageous.

Together, these results establish a general and practical strategy for deep three-photon microscopy, combining adaptive beam shaping with an illumination-matched aberration basis for AO correction. This unified approach enables stable, efficient aberration correction under realistic power constraints and provides a robust foundation for extending high-resolution in vivo imaging to greater depths.

## Acknowledgments

We are grateful to Light Conversion, Photonics Solutions (UK), for kindly lending us the CRONUS-3P laser unit, Bertin Alpao for lending us the DM Alpao-DM97-15, Randy Bruno for providing funding for equipment, Sandra Tan for mouse cranial surgeries, Angus Silver for constructive conversations. This work was supported by a Schmidt Sciences LLC (Schmidt AI in Science postdoctoral research fellowship), Sir Henry Wellcome postdoc fellowship (222807/Z/21/Z), UKRI/Wellcome Physics of Life Grant (EP/W024047/1), and grants from Wellcome (213465/Z/18/Z) and ERC/UKRI (EP/X026655/1).

**Data availability.** Data underlying the results presented in this paper are not publicly available at this time but may be obtained from the authors upon reasonable request.

**Ethics Statement.** All experiments were conducted in accordance with the UK Animals (Scientific Procedures) Act 1986 under appropriate project and personal licences. Adult C57BL/6J mice (male and female, 8–12 weeks old, n = 3) were used. Further details are provided in the Supplementary Methods.

## DEPTH-ADAPTED ADAPTIVE OPTICS FOR THREE-PHOTON MICROSCOPY: SUPPLEMENTAL DOCUMENT

## 1. Depth-optimised illumination beam size: numerical analysis

The beam size optimisation analysis was adapted from the work [1]. We assumed a rotationally symmetrical Gaussian incident field expressed as

$$E(r,\beta) = \mathrm{N}\exp\left(-\frac{r^2}{\beta^2 R^2}\right)$$

where $N$ is a normalisation term defined below, r is the radial coordinate of the incident plane, $\beta$ is the underfilling ratio between the radius of the Gaussian beam and the radius of the objective back aperture and $R$ is the radius of the objective back aperture. If $\beta = 1$, the resultant beam is a truncated Gaussian with a field amplitude at the objective aperture equal to $e^{-1}$, and intensity equal to $e^{-2}$ (or about 13.5%) of the intensity at the centre.

The incident beam is normalised to ensure the total intensity of the incident beam is kept constant while adjusting $\beta$. When integrating the intensity of the Gaussian field between $r = 0$ to $r = R$, the expression can be approximated as

$$\int_0^R E^2(r,\beta)\, r\, dr = \left[C\exp\left(-2\frac{r^2}{\beta^2 R^2}\right)\right]_0^R = C\left(1-\exp\left(-\frac{2}{\beta^2}\right)\right)$$

where $C$ is an integration constant. The normalisation term, $N$, to the total Gaussian incident field integrating over the circular aperture with respect to the field with $\beta = 1$ is thus written as

$$N = \frac{1}{\beta}\sqrt{\frac{1-\exp(-2)}{1-\exp\left(-\frac{2}{\beta^2}\right)}}$$

The incident field can also be expressed in the spherical coordinates of the ray model, written as

$$E(\alpha,\beta) = \mathrm{N}\exp\left(-\frac{\sin^2\alpha}{\beta^2\sin^2\alpha_{\max}}\right)$$

where $\alpha_{max}$ the largest angle of ray incidence, which relates to the NA of the objective lens in the propagation medium. $\alpha$ is the angle of ray incidence. The spherical coordinates expression helps to derive the attenuation term as the light propagates to a distance in tissue.

Light gets attenuated based on the propagation distance (which relates to the imaging depth, $l_{depth}$), and the attenuation depth of the medium, $l_e$, with a relation

$$atten = \exp\left(-\frac{l_{depth}}{2l_e\cos\alpha}\right)$$

Putting the normalisation, Gaussian field and attenuation term together, we can write the incident Gaussian field as

$$E_{gauss}^{inc}(\alpha,\beta) = \frac{1}{\beta}\sqrt{\frac{1-\exp(-2)}{1-\exp\left(-\frac{2}{\beta^2}\right)}}\exp\left(-\frac{\sin^2\alpha}{\beta^2\sin^2\alpha_{max}}\right)\exp\left(-\frac{l_{depth}}{2l_e\cos\alpha}\right)$$

In our analysis, $\alpha_{max}$s determined by the numerical aperture of the objective lens. $l_e$ was the attenuation depth in the mouse cortex and immersion water for our selected illumination wavelength, where we used $l_e = 293\ \mu m$ in the cortex and $l_e = 10\ mm$ in water [2, 3].The working distance of our objective lens was $4\ mm$.

The vectorial electric field model for a high NA objective lens' focal volume at point P (initially presented by Richards and Wolf in [4]) with linear-x-polarised light can be written as (after removing some constant multipliers)

$$E_x(P) = I_0 + I_2\cos 2\phi_P$$
$$E_y(P) = I_2\sin 2\phi_P$$
$$E_z(P) = -2iI_1\sin\phi_P$$

where $\phi_P$ is the angular co-ordinate of point $P$ on the lateral focal plane. $i$ is the imaginary unit. After removing any constant, $I_0$, $I_1$ and $I_2$ at point P can be written as

$$I_0(P) = \int_0^{\alpha_{max}} E_{gauss}^{inc}(\alpha,\beta)\sqrt{\cos\alpha}\sin\alpha\,(1+\cos\alpha)J_0(k\rho\sin\alpha)\exp(ikz\cos\alpha)\,d\alpha$$
$$I_1(P) = \int_0^{\alpha_{max}} E_{gauss}^{inc}(\alpha,\beta)\sqrt{\cos\alpha}\sin^2\alpha\,J_1(k\rho\sin\alpha)\exp(ikz\cos\alpha)\,d\alpha$$
$$I_2(P) = \int_0^{\alpha_{max}} E_{gauss}^{inc}(\alpha,\beta)\sqrt{\cos\alpha}\sin\alpha\,(1-\cos\alpha)J_2(k\rho\sin\alpha)\exp(ikz\cos\alpha)\,d\alpha$$

where $J_m$ is the $m^{th}$ order of Besel functions of the first type, $\rho$ is the radial co-ordinate of point $P$ on the lateral focal plane and $z$ is the axial distance from point $P$ to the focal plane.

And finally, the generated total three-photon signal, $S_3$, for a volume object can be expressed as

$$S_3 = \int_V \left(|E_x(P)|^2 + \left|E_y(P)\right|^2 + |E_z(P)|^2\right)^3 dP$$

where $V$ is the entire focal volume. We calculated numerically the three-photon signal, summing the pixels above the threshold of $1/10^3$ of the maximal value to encompass most of the power of the focus.

For each proposed imaging depth in cortex, we computed and optimised the three-photon signal $S_3$ by varying the filling ratio $\beta$. Results are included in the main paper Figure 1.

## 2. Beam profile measurement

To assess the effect of the illumination underfilling ratio, we characterised the imaging system by measuring the beam profile and power at the objective back aperture while adjusting the beam expander to adjust the beam size, and the half-waveplate to adjust the beam power.

For beam-profile measurements, the beam expander setting was fixed at infinite conjugate to ensure a collimated output beam. The expansion ratio was varied from 0.5 × to 2.5 ×. The beam profile was measured using a power meter with a circular aperture of $9.5mm$ (Thorlabs PM100D, SC405C sensor head). To obtain the intensity distribution, black tape was applied to the power-meter head, leaving a narrow central slit ($\sim 1mm$) exposed. The taped power meter was mounted near the objective back aperture and translated laterally in 1 mm steps; power was recorded at each position.

The measured slit-integrated power was used to reconstruct the beam profile by assuming a rotationally symmetric Gaussian shape, consistent with the manufacturer's manual and visual inspection using an IR card and viewer. In addition, a point-based 2-D beam scan was also used to verify the Gaussian beam size and beam shape.  A The underfilling ratio, β, was calculated as the ratio of the beam diameter (cut-off at 13.5% of peak intensity) to the objective aperture diameter.

Beam power after the objective was also measured for each expansion setting while adjusting the half-waveplate, ensuring fair comparison at constant illumination power across imaging depths. This calibration process was performed once, and the beam expander was found to increase beam size linearly and consistently within measurement uncertainty.

## 3. Three-photon fluorescence microscopy

Three-photon fluorescence imaging was carried out using a Cronus 3P laser system (Light Conversion), which integrates a tuneable optical parametric amplifier and prism compressor, powered by a 1030 nm femtosecond laser. The setup was optimized for 1300 nm excitation and delivered 49 fs pulses (measured under the objective using an APE autocorrelator with an external detector) with a 1 MHz repetition rate. During animal experiments, the laser power was adjusted to 50 mW at the sample. Image acquisition was conducted using a galvanometer-based multiphoton system (VivoScope, Scientifica, UK) equipped with a water-immersion objective (Olympus, XLPLN25XSVMP2, 25×/1.0 NA, 4 mm WD).

Green and red fluorescence signals were detected via separate PMT channels (Hamamatsu H10770PA-40) using dichroic (T565LPXR, Chroma) and bandpass filters (ET525/50m-2p for green, ET620/60m for red, Chroma). The signals were processed through a 150 kOhm, 1 MHz transimpedance amplifier (XPG-ADC-PREAMP) and digitized using a vDAQ card (Vidrio Technologies). ScanImage 2021 software (Vidrio Technologies), running on MATLAB, managed the image acquisition. The system included a motorized stage and focusing unit to control sample and objective positioning. Imaging depths were measured via axial displacement of the stage. For most structural scans, we used a 256×256 resolution with 3.2 μs dwell time at 4.23 Hz, averaging across 240 frames unless noted otherwise.

## 4. Adaptive optics and aberration correction

### 4.1 Deformable mirror (DM) calibration

Prior to usage, the continuous-membrane DM underwent a complete Zernike mode calibration using a custom-built interferometer [5]. Each actuator was systematically driven across voltages while interferograms were captured. These were processed via custom Python software (https://github.com/dop-oxford/dmlib) to extract wavefronts, fit actuator responses to Zernike basis functions, and generate a mode-to-voltage transformation matrix. This ensured accurate rendering of any target wavefront correction during imaging.

### 4.2 Optical relay path

To compensate for optical distortions during deep tissue imaging, one of two DMs (Imagine Optic Mirao 52e or Alpao DM97) was alternatively introduced into the beam path. The beam, expanded post-OPA to match the DM's aperture (15 mm and 13.5 mm respectively), was directed through a telescope system that aligned the DM with both the scanning mirrors and the back pupil plane of the objective, ensuring uniform correction across the field of view, via two relay-lens pairs of 500/60 mm and 30/206 mm focal lengths (not shown in Figure 1a). A secondary 8.3× beam reduction with a telescope with lenses of 500 mm and 60 mm focal lengths followed, bringing the beam size to 1.8 mm (1.62 mm for DM97) at the Galvo scanners, which was then re-expanded at the objective pupil to 12.75 mm (11.57 mm for DM97), resulting in a diameter fill ratio (DM aperture to objective aperture) of 0.85.

4.3 Sensorless AO correction

During imaging experiments, we used a sensorless AO approach through which image quality metrics—such as brightness or contrast—served as feedback to guide correction. We iteratively applied varying amplitudes of specific Zernike modes and measured image quality metrics to identify the optimal correction. Tip, tilt, and defocus modes were excluded to avoid field-of-view displacement. The remaining modes were optimized one at a time by fitting a parabola or Gaussian curve to metric values across five evenly spaced bias RMS amplitudes within the range of ±1.5 or ±3 rad (as stated in each set of results). This Gaussian fitting enabled estimation of the best correction within the selected bias range for each aberration.

We employed the 5N correction scheme, applying five biases per mode. For animal corrections, primary astigmatism, coma and trefoil in Zernike (Z_2^(±2), Z_3^(±1), Z_3^(±3)) and their corresponding w-Bessel modes (B_2^(±2), B_3^(±1), B_3^(±3)) were included. Though originally demonstrated on two-photon systems [6], the methodology extends naturally to three-photon excitation [7].

For system flat DM pattern generation, the objective correction collar was first adjusted to compensate for coverslip-induced spherical aberration. Based on the DM flat pattern, the system flat DM pattern was measured from a beads sample using standard sensorless AO algorithm when the beam was adjusted to $\beta \approx 0.72$, which served as a starting point for subsequent correction. Seven Zernike modes (Z_2^(±2), Z_3^(±1), Z_3^(±3), Z_4^0) were included in the correction.

## 5. Animal preparation

All animal work adhered to the UK Animals (Scientific Procedures) Act 1986. Experiments were carried out using adult C57BL/6 mice, including both males and females, aged between 2 and 6 months. The study included three genetically modified mice from the CamKIIα–tTa (AI94) × B6; DBA-Tg(tetO-GCaMP) line, with animals expressing GCaMP8s. Mice were housed in controlled conditions, maintained at 20–22 °C with 40% relative humidity, on a standard 12-hour light/dark cycle. Animals had unrestricted access to food, and their health and wellbeing were monitored daily.

Animal surgeries were performed following the same protocol described in [8]. Briefly, anaesthesia was induced with 3% isoflurane and maintained at or below 1.5% throughout the procedure. Mice received perioperative analgesia with subcutaneous buprenorphine and meloxicam, and the scalp was locally treated with bupivacaine prior to sterilisation and incision. After cleaning the skull and attaching a custom aluminium head-plate, a circular craniotomy (4–5 mm diameter) targeting the prelimbic cortex was performed. For animals receiving viral labelling, a retrograde virus (rAAV2-CAG-tdTomato) was injected into the dorsomedial striatum (DMS) using a glass pipette and nanoinjector, followed by durotomy and cranial window implantation with a double coverslip. Mice were allowed to recover in a heated chamber and monitored for 7 days, with a minimum recovery period of 21 days prior to behavioural experiments to allow for viral expression.

For the beads slide, 0.5 μm red fluorescent beads (Invitrogen™ # F8812) were diluted 100 times before being laid on standard #1.5 microscope coverslip to dry. VECTASHIELD Antifade mounting medium (Vector Laboratories, UK) was used to mount the beads-coated coverslips on microscope slides. They were then sealed with nail vanish.

## 6. Modal comparison and analysis

To further examine the stability of different aberration bases, we compared the effects of the Zernike coma modes ($Z_3^{\pm1}$) and their weighted-Bessel counterparts ($B_3^{\pm1}$) using fluorescent bead samples (Supplementary Figure 4). When applied over a range of phase amplitudes, the Zernike coma modes produced substantial modulation of fluorescence intensity, accompanied by noticeable lateral shifts in the bead centre of gravity in the x–z plane. In contrast, the corresponding w-Bessel modes yielded comparable intensity modulation while substantially reducing lateral displacement, indicating improved decoupling between aberration correction and beam translations.

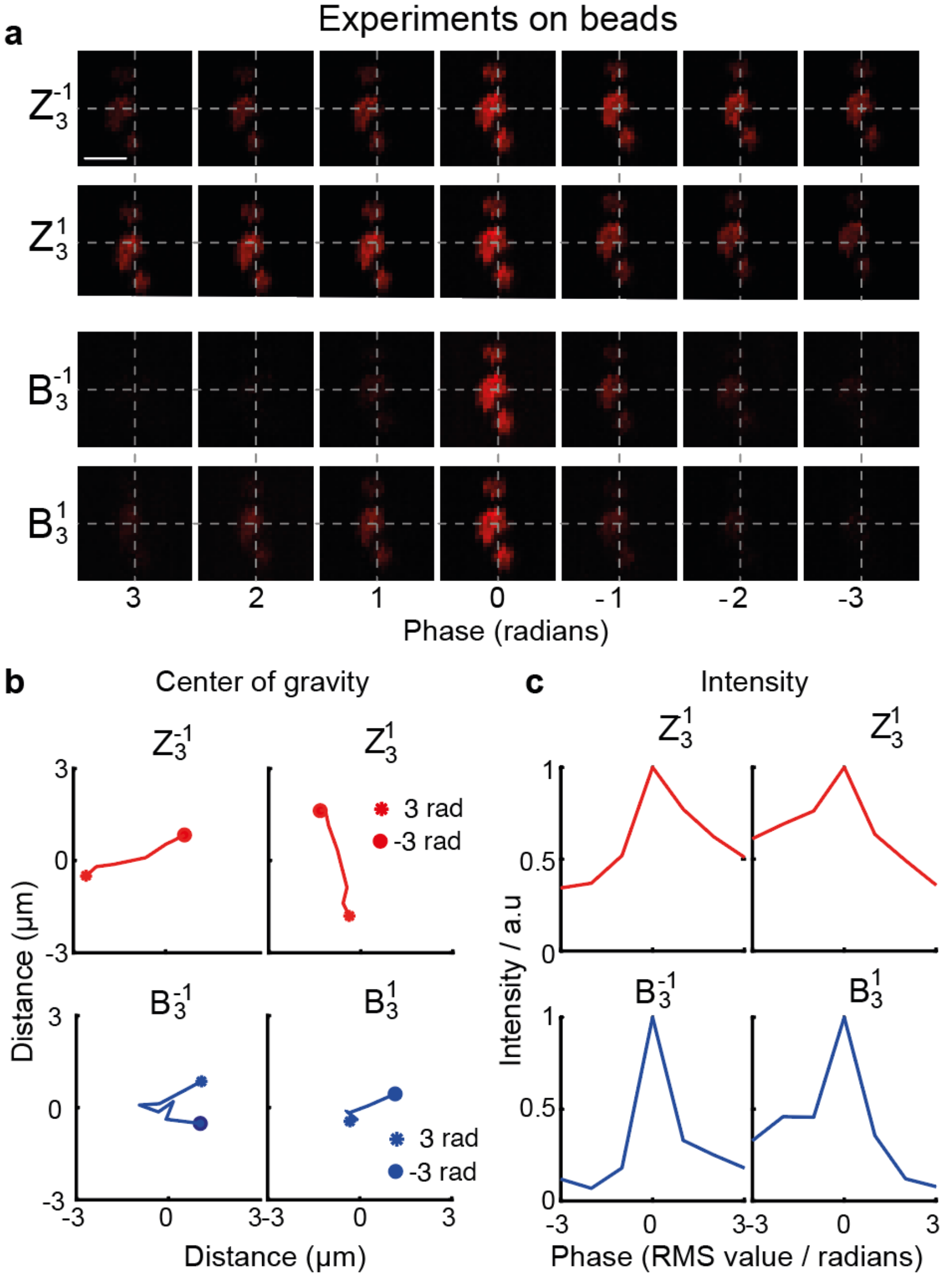

Fig. S1. **Field of view shifts induced by Zernike but not by weighted-Bessel aberration modes.** (a) x–z bead images acquired while applying Zernike and weighted-Bessel modes across phase amplitudes from −3 to +3 rad. (b) Lateral centre-of-gravity displacement as a function of phase for Zernike (red) and weighted-Bessel (blue) modes. (c) Normalised fluorescence intensity as a function of phase for the same modes.

For comparison, we also examined the effect of the Zernike spherical mode ($Z_0^4$) and its corresponding w-Bessel mode ($B_0^4$) in bead measurements (Supplementary Figure 2). In this case, the applied spherical aberration resulted in relatively weak intensity modulation but induced a pronounced axial shift of the focal plane. This behaviour reflects strong cross-talk between spherical aberration and defocus under truncated Gaussian illumination, consistent with the non-orthogonality observed in the Zernike basis. Together, these results highlight that w-Bessel modes provide improved control of aberrations while minimising unintended focal shifts, which is critical for stable sensorless adaptive optics correction.

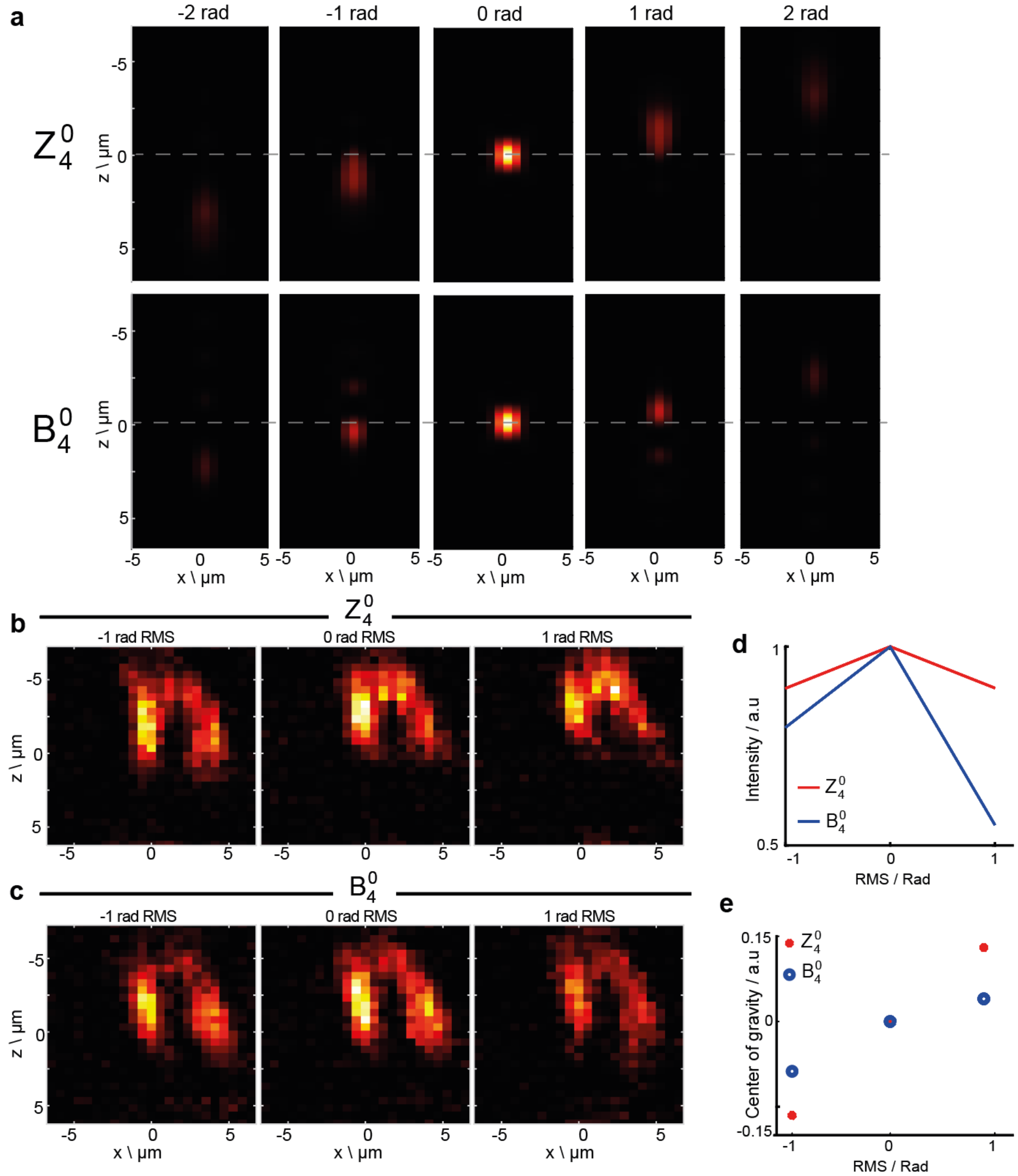

Fig. S2. **Differential effects of Zernike and weighted-Bessel spherical modes on focus quality and axial stability.** (a) Simulation for Zernike and Bessel spherical aberration modes. (b) x–z plane images of a fluorescent bead sample obtained while applying the Zernike spherical aberration mode $Z_4^0$ with amplitudes of −1, 0, and +1 rad RMS (left to right). (c) Weighted-Bessel $B_4^0$ mode applied at −1, 0, and +1 rad RMS. (d) Corresponding normalised fluorescence intensity as a function of applied RMS phase for $Z_4^0$ (red) and weighted-Bessel $B_4^0$

(blue) modes. (e) Corresponding axial centre-of-gravity shifts of the bead signal as a function of applied RMS phase for Zernike $Z_4^0$ (red) and weighted-Bessel $B_4^0$ (blue).

## 7. Additional experimental results on beam expansion effects

Additional experimental results demonstrating the effects of beam expansion on imaging quality at different cortical depths are shown in Supplementary Figure 3. Similar depth-dependent trends were observed across multiple animals and imaging sessions, and the optimal filling ratios were consistent across different illumination power levels, indicating that the effect was robust and not dependent on a specific preparation or power regime.

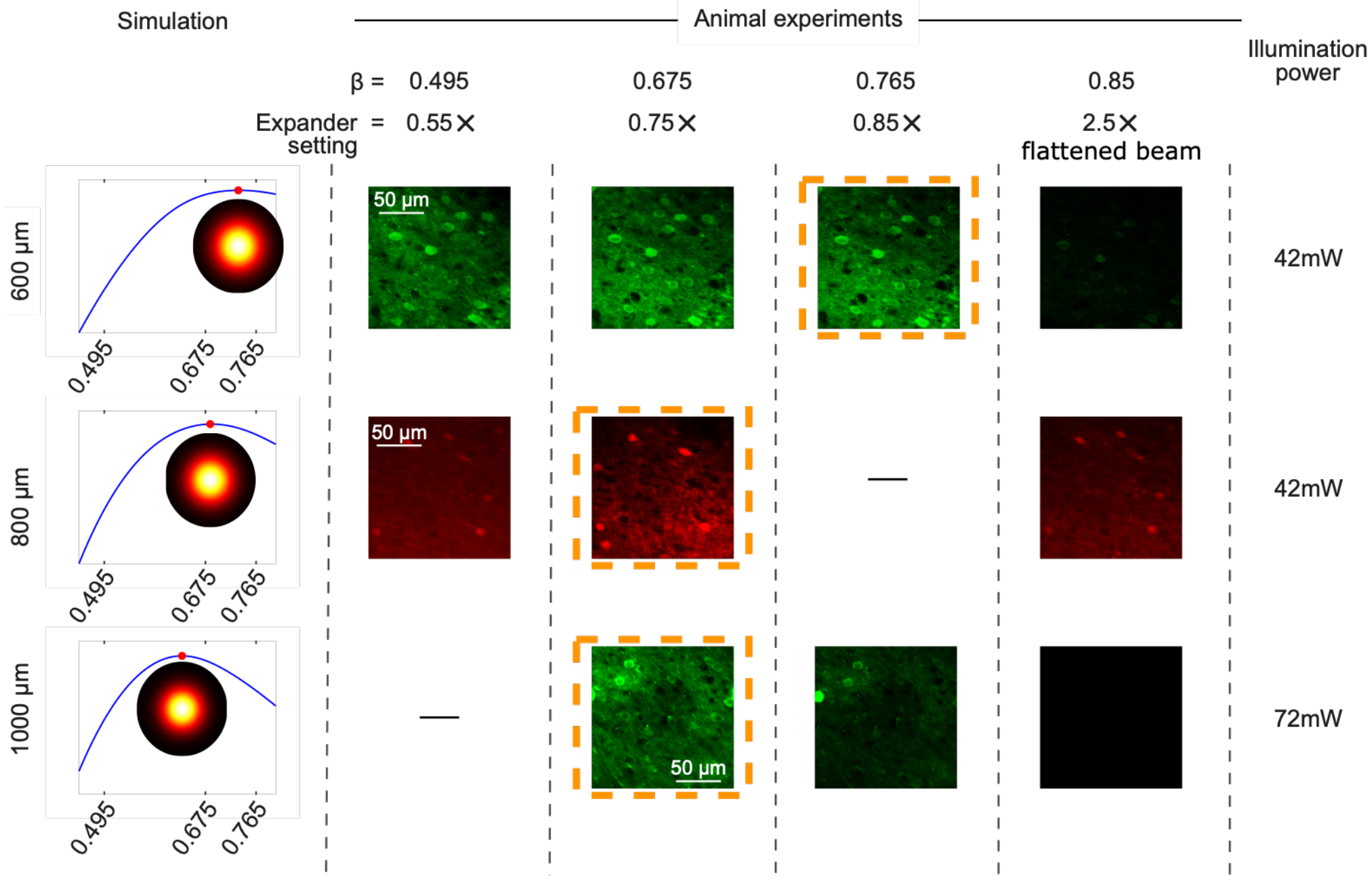

Fig. S3. **Depth-dependent optimisation of illumination beam expansion.** Simulation of three-photon signal as a function of illumination filling ratio (β) at 600, 800, and 1000 µm depth, with corresponding in vivo images acquired using different beam-expander settings and a flattened beam profile. At each depth, the brightest images correspond to the β values predicted by simulation.

## 8. Experimental results on modal comparison

Additional experimental results comparing Zernike modes and w-Bessel modes were included in supplementary Figure 4. Two experiments at 900 $\mu m$ and 1000 $\mu m$ depth was shown. W-Bessel modes showed consistently superior corrections as evidenced through the resultant images being brighter and sharper, with a larger number of high pixel values. The correction also converged faster, where w-Bessel generally optimised after the first iteration, and Zernike modes had coefficients fluctuating even after three iterations of correction.

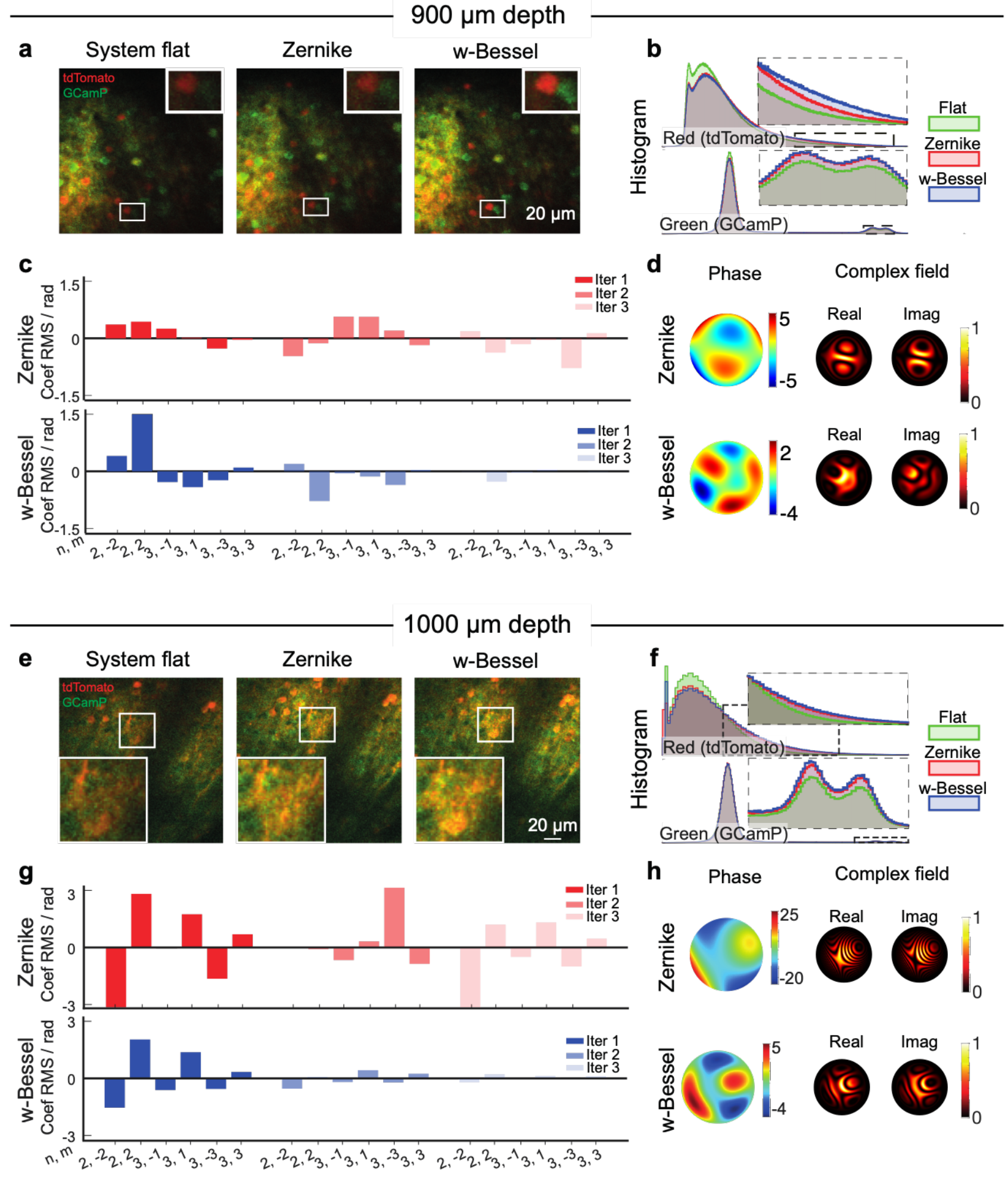

Fig. S4. **Comparisons of w-Bessel and Zernike mode based correction of aberrations** at (a-d) 900 $\mu m$ and (e-h) 1000 $\mu m$ in an awake mouse cortex for the designed beam profile. (a, e) The selected FOV averaged from 120 frames when the AO was applying a system correction (left image), and after applying corrections generated from the Zernike set (middle image) and w-Bessel set (right image). (b, f) the histogram comparison of the images before and after corrections. (c, g)The bar charts showed the computed modal coefficients after each of the three iterations of corrections using Zernike and w-Bessel modes. (d, h) The corresponding wavefront corrections and the resultant gaussian illumination with real and imaginary intensity components.